\documentstyle[aps,prl,graphicx]{revtex}

\begin{document}
\title{Possible indication of narrow
baryonic resonances produced in the 1720 - 1790 MeV mass region}

\author{
B. Tatischeff$^{1}$\thanks{E-mail : tati@ipno.in2p3.fr},
J. Yonnet$^{1}$,
M. Boivin$^{2}$, M. P. Comets$^{1}$, P. Courtat$^{1}$,\\
R. Gacougnolle$^{1}$, Y. Le Bornec$^{1}$,
E. Loireleux$^{1}$, M. MacCormick$^{1}$, \\
F. Reide$^{1}$, and N. Willis$^{1}$}

\address{
$^1$Institut de Physique Nucl\'eaire,CNRS/IN2P3, F--91406 Orsay Cedex,
France\\
$^2$Laboratoire National Saturne,CNRS/IN2P3, F--91191
Gif-sur-Yvette Cedex, France}
	  
\maketitle
\vspace*{1cm}
\begin{abstract}
Signals of two narrow structures at M=1747 MeV and 1772 MeV were
observed in the invariant masses M$_{pX}$ and M$_{\pi^{+}X}$
of the  pp$\rightarrow$ppX and pp$\rightarrow$p$\pi^{+}$X reactions
respectively. Many tests were made to see if these structures could
have been produced
by experimental artefacts. Their small widths and the stability of the
extracted masses lead us to conclude that these structures
are genuine and
may correspond to new exotic baryons. Several attempts to
identify them,
including the possible `missing baryons'' approach, are discussed.
\end{abstract}


\section{INTRODUCTION}
Many baryonic states were predicted to exist
above 1.5 GeV, first by
the non-relativistic QCD inspired quark models \cite{smit}, but were
unobserved in $\pi$N elastic scattering.
They are the so-called
`missing baryons''. Molecules between baryons
and mesons and also gluonic states may exist. Low mass, narrow
exotic baryons
have already been observed at Saturne \cite{bor2}, the first was in the
missing mass of the pp$\rightarrow$p$\pi^{+}$X reaction
between the N and the $\Delta$(1232).
Weaker signatures of other narrow baryonic structures have also been
observed in different mass ranges \cite{bor3}, but due to their
weakness, remain unconfirmed. Structures were also observed at
the Moscow Linear Accelerator in the pd$\to$ppX reaction \cite{filk}.
These structures were tentatively
associated with colored quark clusters. A recent paper
describes, in detail, the experimental set up, the analysis, the checks
performed, and the results observed in the baryonic mass range below
1.46 GeV \cite{bor5}.\\
\hspace*{4.mm}The results obtained in the higher baryonic mass range
above 1.5 GeV do not appear to follow an obvious pattern, and
so appear to be somewhat chaotic. These results will not be presented
or discussed in this paper.
Indeed, this mass range corresponds to a region where many
broad and widely excited
$\Delta$ and N$^{*}$ resonances exist and are documented in the
Particle Data Group (PDG) records \cite{pdg}. In this region many
observations can result
from interferences between different broad resonances, having the same
quantum numbers and which overlap. However, considering the two, three
and four star
broad baryonic resonances as quoted by the Particle Data Group \cite{pdg},
the mass range 1720$\le$N$\le$1900~MeV is free of any broad
resonance and is therefore a more appropriate region for the present studies.
It is
important to search for narrow baryonic structures in this mass range since
it can shed light on the open question of the missing baryons. The
measurements
were performed at several incident proton energies \cite{bor2} \cite{bor3}
\cite{bor5}. The higher
energy T$_{p}$=2.1~GeV, gives full access to the interesting region of the
invariant masses M$_{pX}$ and M$_{\pi^{+}X}$ of the  pp$\rightarrow$ppX and
pp$\rightarrow$p$\pi^{+}$X reactions respectively.
\section{Experiment}
\subsection{Experimental set-up}
\hspace*{0.4cm}An experiment aimed to look for such exotic states
was performed using the Saturne (T$_{p}$=2.1 GeV) polarized proton beam
and the SPES3 spectrometer. Figure 1 shows the experimental layout.
Briefly, the main properties of the SPES3 spectrometer are the following:\\
\hspace*{0.2cm}- it is a mean value solid angle
spectrometer ($\pm$50 mrd in both the horizontal and vertical planes), and\\
\hspace*{0.2cm}- it is a large momentum range spectrometer
600$\le$ pc$\le$ 1400 MeV.\\
\begin{center}
\begin{figure}
\vspace*{-0.8cm}
\scalebox{.85}[.73]{
\includegraphics[bb=10 150 580 700,clip=]{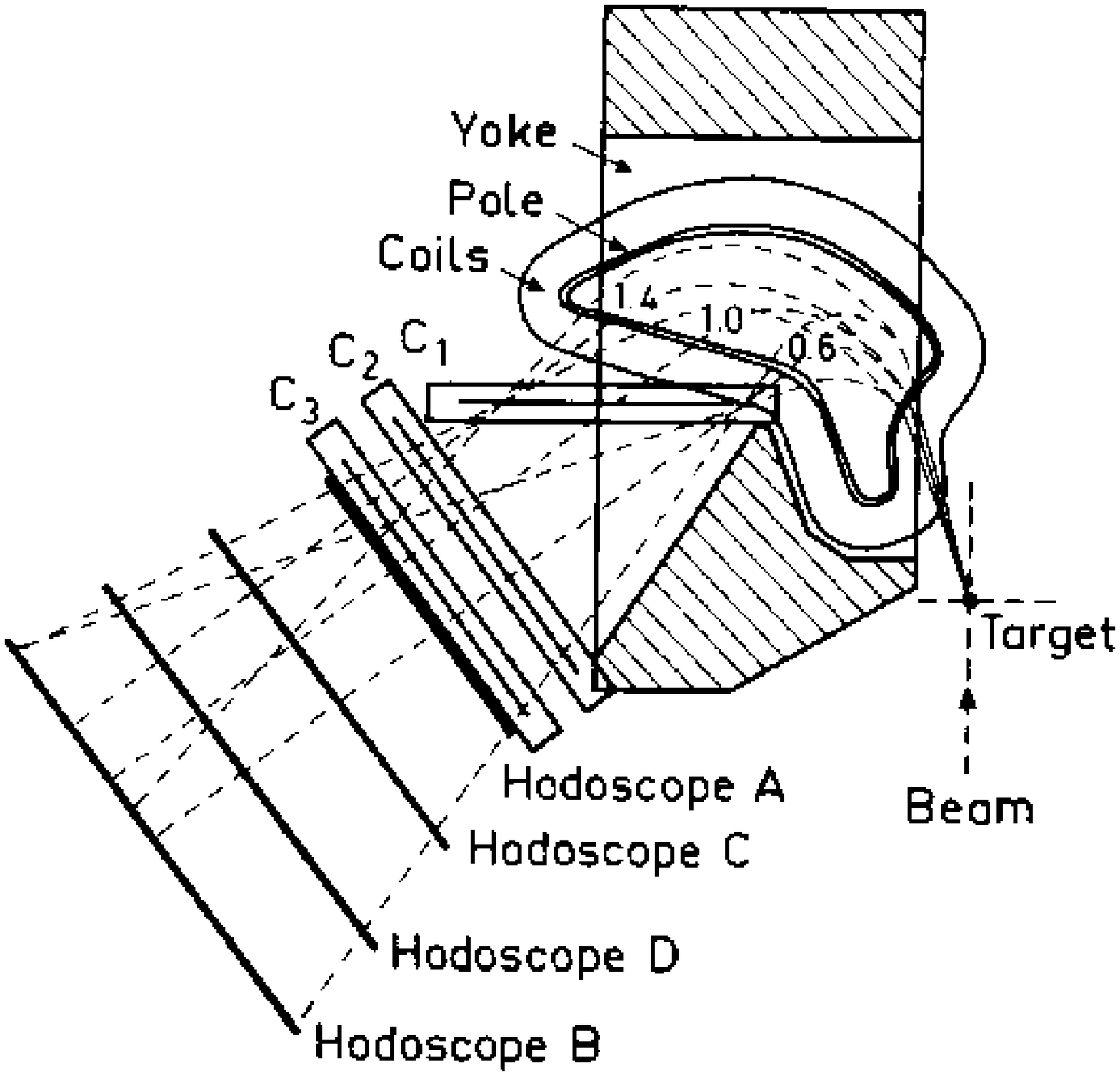}}
\caption{The SPES3 spectrometer and the associated
detection system.}
\label{fig1}
\end{figure}
\end{center}
\hspace*{0.4cm}The experimental conditions were described in detail in a
previous paper \cite{bor5} and will not be repeated here. Its main
properties are
summarized in the following paragraphs.\\
\hspace*{0.4cm}The broad momentum acceptance
allowed the simultaneous study of a large
range of proton and pion energies. The protons were incident on a
liquid hydrogen target
(393 mg/cm$^{2}$ thick). The beam varied between 10$^{8}$ particles/burst
and 5x10$^{8}$ particles/burst, depending on the spectrometer angle,
with the acquisition dead time being less than 10\%. The duration of each
burst
depended on the incident proton energy, and was typically
300 ms for T$_{p}$=2.1 GeV.\\
\hspace*{0.4cm}The particle trajectories were determined
using drift chambers.  The first chamber C1 (`MIT-type''), was situated on
the spectrometer focal plane. Its horizontal spatial and angular
resolutions were : $\sigma_x$~=~90 $\mu$m and
$\sigma_{\theta}$~=~18~mrd respectively. Two multidrift chambers noted as
C2-C3 in Figure~1, or `CERN chambers'', placed perpendicular to the mean
particle direction, were designed to obtain
trajectories in the horizontal and vertical planes.
However, due to the small vertical magnification of the spectrometer
($\approx$~.14) the $\phi$ resolution at the target was poor. The CERN
chambers were used to determine the
MIT chamber efficiency
by calculating the ratio of three-hit to two-hit coincidences. During the
experiment
the maximum value of the MIT chamber efficiency was 96$\%$ and the variation
of this efficiency was monotonous and continuous along the focal plane
\cite{bor5}.\\
\hspace*{0.4cm} The trigger consisted of four planes of plastic scintillator
hodoscopes. The dimensions of each plastic detector were 12$\times$40~cm$^2$
for the first plane (A), and
18$\times$80~cm$^2$ for the last plane (B). Each of these two planes involved
20 scintillators.  The
time of flight baseline from the first scintillator plane to the last
one was 3~m. Particles were identified by their time of flight
between the A$_i$ and B$_j$ detectors and also by their energy loss
in the A$_i$ plane. This latter measurement was mainly used
to discriminate between one and two charged particles.
The large horizontal angular magnification of the spectrometer
resulted in a wide angular opening (up to $30^0$).
It resulted in a large number of possible A$_i$.B$_j$ combinations (125),
between the first and
last scintillator counter planes. It is important to note that a
mean range of 200 MeV/c ($25\%$
of the focal plane acceptance) is covered by each A$_i$.B$_j$ combination.
There is therefore a large overlap between many A$_i$.B$_j$ trigger
combinations for each spectrometer momentum.
Careful calibrations and efficiency measurements of all the 125 combinations
were performed using a system of scintillators that could be automatically
displaced. This system was installed in front
of the `A''-hodoscope and behind the `B''-hodoscope. The trigger efficiency
mean value was about $95\%$.\\
\hspace*{0.4cm}Two particles - either two protons, or a proton
and a pion - were identified in the final state using two times of
flight. The
first time was measured behind the spectrometer, over a 3 m baseline,
and allowed a very good
separation between p and $\pi^{+}$, since each scintillator had an intrinsic
resolution of $\sigma$=180 ps. The second time
measurement was made in the `A''
plane of scintillators, for the two detected particles over the 6 - 7 m
distance from the target. This provided the means to control the random
coincidence contribution and to reject the small amount of badly identified
reactions. A correction was made to
take into account the differences in trajectory lengths, and then a common
window of $\pm$~2~ns was used for all the 190 (19x20/2) time of flight
channels. The resolution of this distribution
is $\sigma~\approx$~570 ps.
In the analysis, roughly $0.6\%$ of events were eliminated due to a timing
mismatch between the trigger and chamber data.\\
\hspace*{0.4cm}These studies require good resolutions and
good statistics for the results to be studied in a sufficiently large
number of data bins. This was achieved in this
experiment. The typical resolution was
$\sigma(M)~\approx$~3~MeV
(4~MeV) for $\theta$=3$^{0}$ (9$^{0}$), and the acquired statistics
amounted to 1500 counts (after software cuts) per 1 MeV
invariant mass bins.\\
\subsection{Search for experimental artefacts}
\hspace*{0.4cm}All elements of the detection system were
calibrated. As well as this, the effect of possible inefficient or
hot wires in the MIT chamber was studied in detail. All final data
(missing masses or invariant masses) were the results of two-particle
coincidences.
Any possible inefficient or hot wire in the MIT chamber would affect both
particles indifferently. The consequence would be a weak  - or intense  -
line over a narrow range in the momentum scatterplots for both detected
particles. Given that this paper is devoted to the search for narrow
baryonic resonance structures and that this kind of experimental defect
could effectively simulate such an effect, it is important to show that this
eventuality has been explored in detail.\\
Each momentum is reconstructed using several (from 3 up to 7) wires in the
MIT chamber. The number of wires depends on the trajectory angle.
One inefficient wire shifts the reconstructed momenta by an amount
less than or equal to one half interwire distance, that is $\le$~1.14~MeV/c.
Since each mass is determined using a combination of large momenta ranges for
both detected particles (see figure 2), the effect of one inefficient wire
must be small. We illustrate such effect in figure 3. Figure 3(a) shows the
\begin{center}
\begin{figure}[h]
\scalebox{.85}[.63]{
\includegraphics[bb=5 5 530 530,clip=]{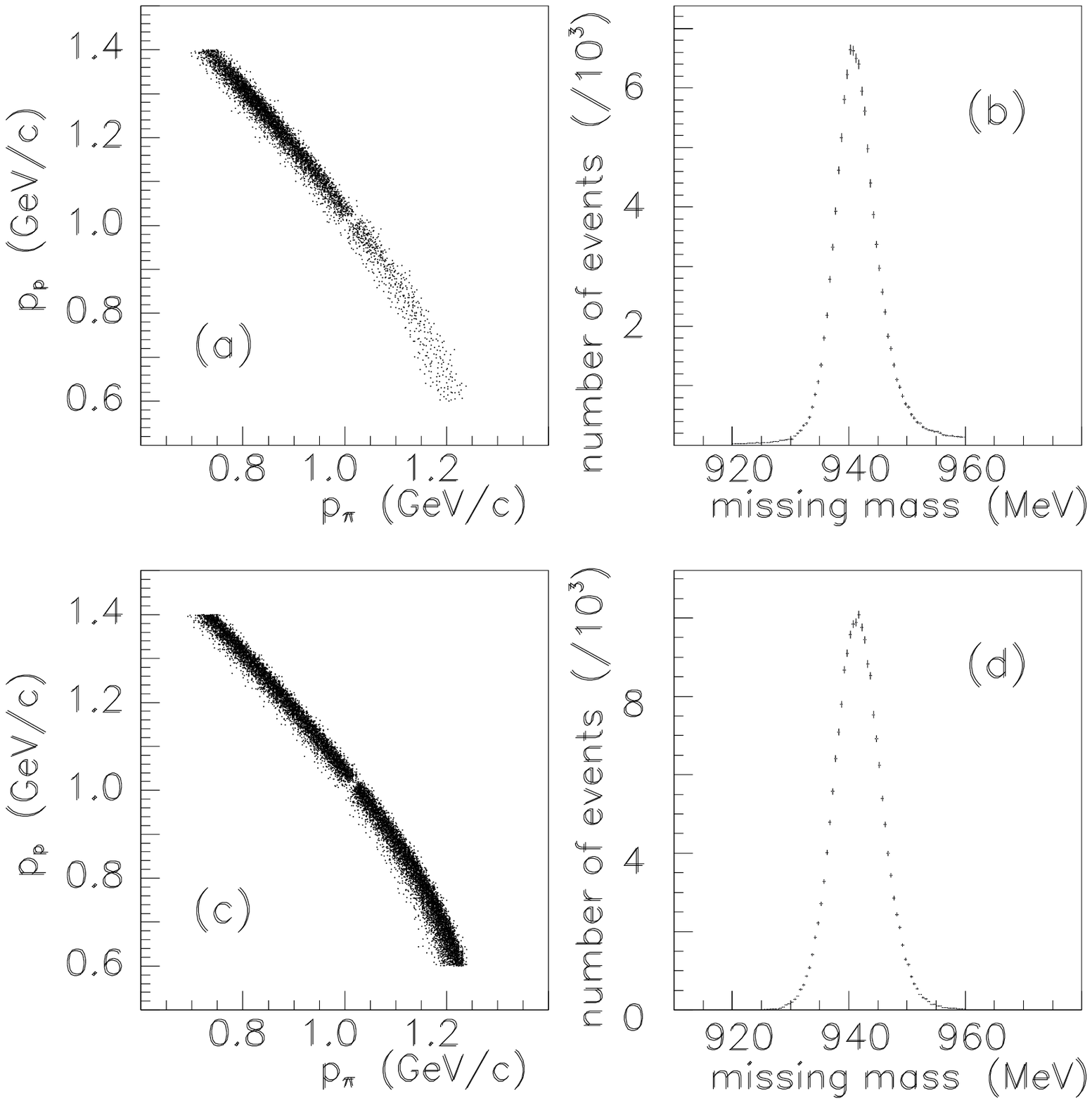}}
\caption{The pp$\rightarrow$p$\pi^{+}$n reaction at
$T_{p}$~=~1520 MeV, $\theta$~=~2$^{0}$. Comparison between data: inserts (a)
and (b), and simulation: inserts (c) and (d).}
\label{fig2}
\end{figure}

\begin{figure}[h]
\vspace*{-0.8cm}
\scalebox{.85}[.53]{
\includegraphics[bb=5 5 530 530,clip=]{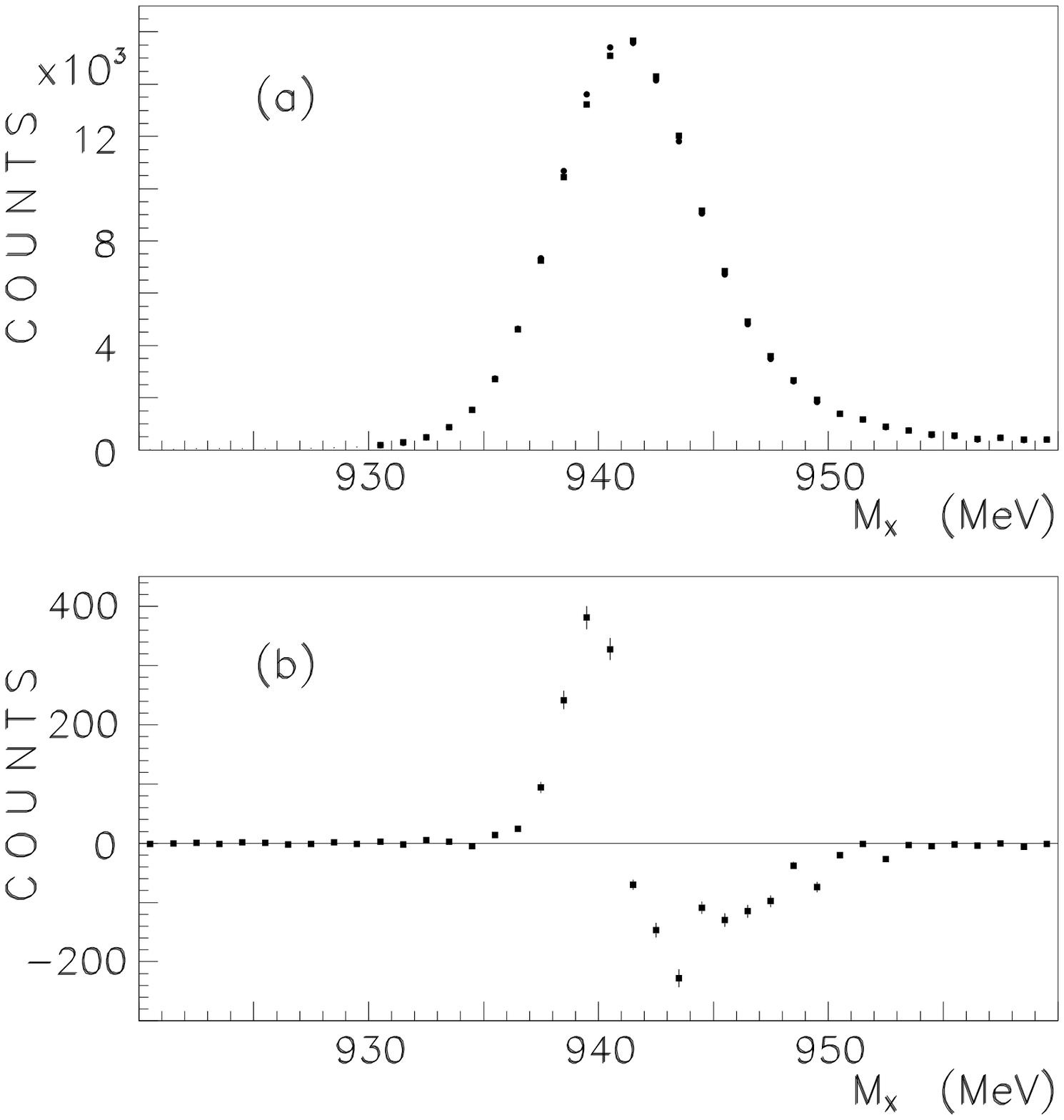}}
\caption{Study of the effect of an eventual
inefficient wire in the MIT drift chamber (see text). Full squares
correspond to the original analysis whereas full circles correspond
to partially shifted data.}
\label{fig3}
\end{figure}
\end{center}
 Figure 3(a) shows the
missing mass at T$_{p}$=1520~MeV, $\theta$=2$^{0}$, and also the same missing
mass when a shift of 1.14~MeV/c is performed on both detected momenta in
the range 900$\le$~p~$\le$~915~MeV/c. The applied shift is larger
than the real one, since the shift of the momentum is smaller than 1.14 MeV/c
when the number of useful wires increases.
For such shift, we obtain a missing
mass distribution which is difficult to discern from the previous one.
Figure 3(b) shows the difference between both missing mass spectra.
When applying the inefficient wire effect described previously, the
resulting neutron mass, and the $\sigma$ of the gaussian describing
the distribution, move by a negligible amount. In the
previous example,
$\Delta~M\approx$~4~keV and $\Delta\sigma\approx$~1~keV respectively.\\
\subsection{Analysis}
The data were normalized using the inefficiency calibrations. They were
also normalized using inefficiencies and
acceptances obtained using the results of a simulation code which is
described in section ``Simulation''.
The data were normalized by the useful momenta ranges of the two detected
particles. The correction function varies as a function of the invariant mass
under study and consequently, the final error bars are dominated by the
statistics and are not only related to the cross section values.\\
\hspace*{0.4cm}
During the raw data analysis, several software cuts were applied
at the expense of the statistics but for the benefit of a more stringent
selection. The quadratic kinematical equations
allow two roots for the invariant
baryonic masses studied in this work. The corresponding events must be studied
separately for the following reasons:\\
- due to the momentum limits of the spectrometer, the limits of the
invariant masses for both root values are different. A simple sum of both
spectra would have created a large shoulder in the result,\\
- the normalization by the momentum acceptance $\Delta$p$_{1}.\Delta$p$_{2}$
is different in each case, and,\\
- the empty region between the two filled
regions is not constant but decreases for increasing masses and tends to zero
as the kinematics tends to the double value limit.\\
\hspace*{0.4cm}The events corresponding to both
root values were separated
by  selecting particular proton momenta for the pp$\rightarrow$p$\pi^{+}$X
reaction. We will call upper (lower) branch spectra the spectra built from
events with momenta larger (smaller) than the upper mass limit momenta.
During the study of the
pp$\rightarrow$ppX reaction the two protons were classified as slow
or fast momenta protons according to their relative values.\\
\hspace*{0.4cm} These two reactions are shown in figures 4, 5, and 6 which
display the differential cross sections versus the
following invariant masses: either
$M_{p(slow)X}\equiv M_{p_{s} X}$, or
$M_{p(fast)X}\equiv M_{p_{f} X}$.
Both roots were separated by software cuts keeping forward or backward
missing mass C.M. angles. These software cuts were applied to the momenta
of the particle not used in the invariant mass combination.\\
\begin{center}
\begin{figure}[h]
\scalebox{.85}[.63]{
\includegraphics[bb=25 350 500 780,clip=]{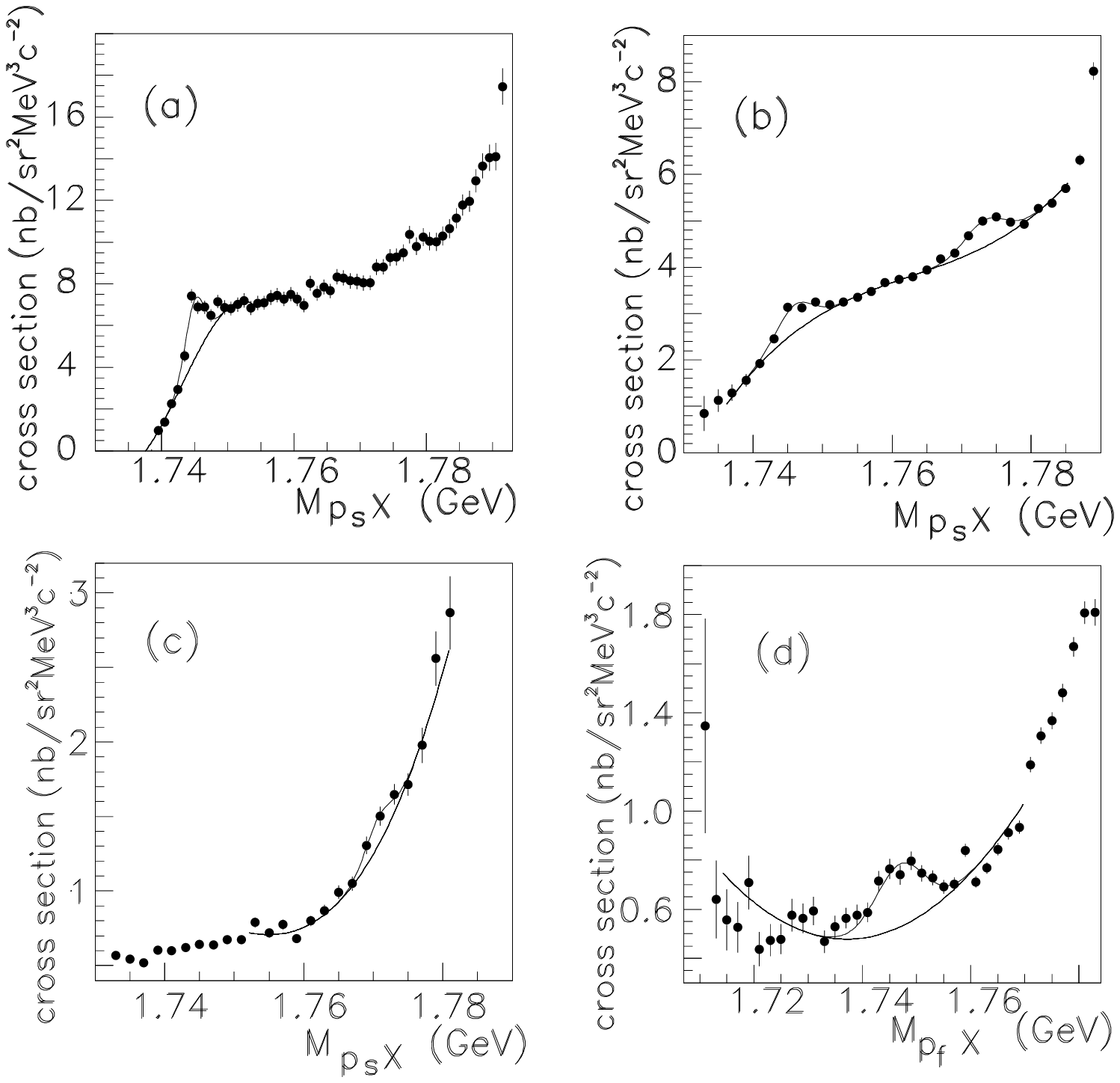}}
\caption{Differential cross sections for the reaction
p~p~$\to$~p$_{f}$~p$_{s}$~X at T$_{p}$=2.1 GeV, in the laboratory system.
Inserts (a), (b), and (c) show the cross sections versus M$_{p_{s}X}$ at
the following angles $\theta$=0.7$^{0}$, 3$^{0}$, and 9$^{0}$ respectively.
Insert (d) shows the cross sections of the upper branch versus
M$_{p_{f}X}$ at $\theta$=9$^{0}$. Values are listed in Table~1.}
\label{fig4}
\end{figure}
\end{center}
\hspace*{0.4cm}For both reactions, if we write it as
$p_{1}\times$$p_{2}\to p_{3}\times$$M_{inv.}$, then the two roots are
separated by the p$_{3}$ momentum and the upper (lower) branch is also
the forward (backward) solution for the two particle-reaction.
\begin{center}
\begin{figure}[h]
\scalebox{.85}[.63]{
\includegraphics[bb=70 350 530 770,clip=]{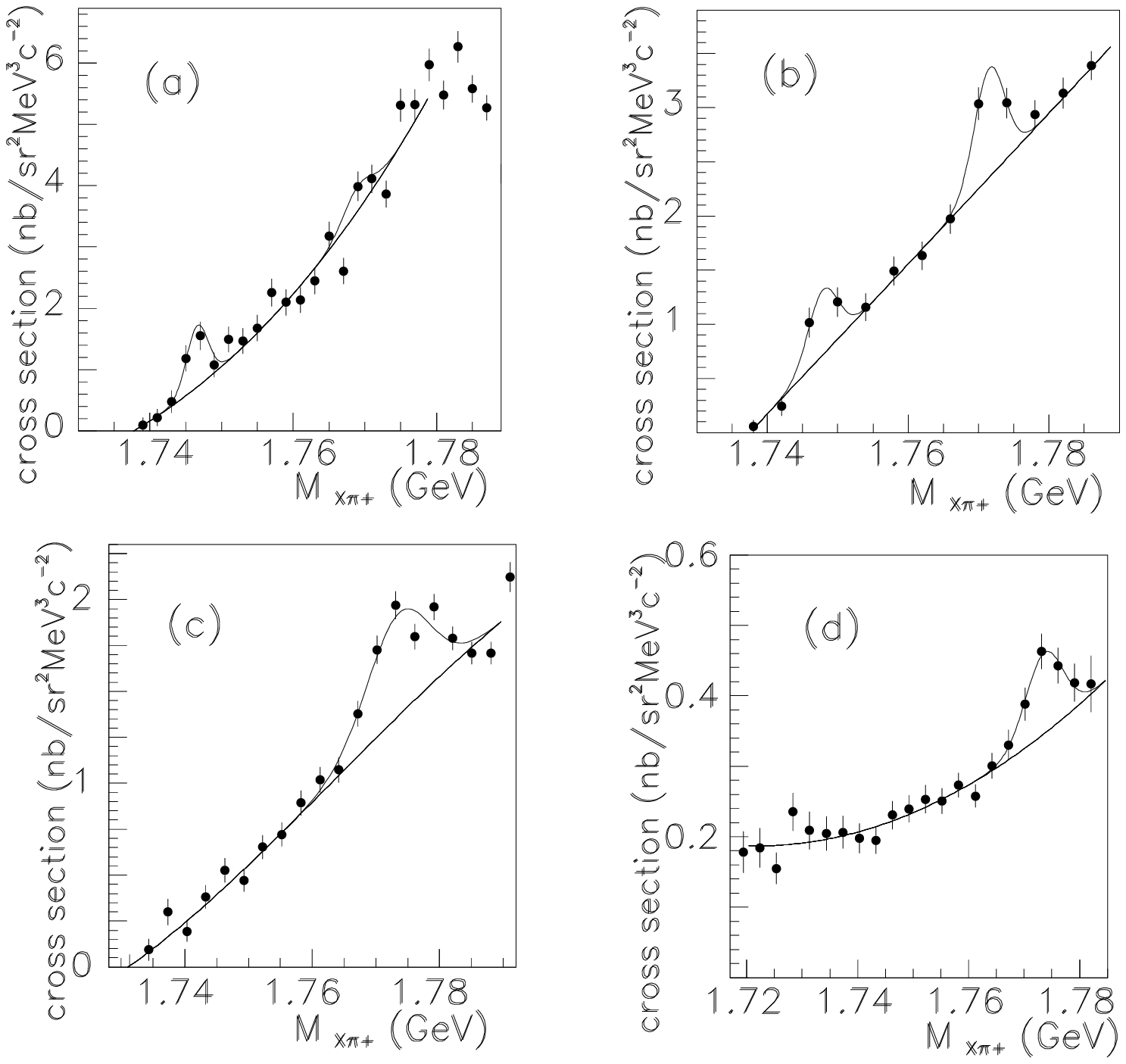}}
\caption{Differential cross sections for the reaction
p~p~$\to$~p~$\pi^{+}$~X at T$_{p}$=2.1 GeV, in the laboratory
system. Inserts (a), (b), (c), and (d) correspond respectively to
$\theta$=0.7$^{0}$, 0.7$^{0}$ (with different software cuts), 3$^{0}$,
and 9$^{0}$.
Values are listed in Table~1.}
\label{fig5}
\end{figure}

\begin{figure}
\scalebox{.85}[.63]{
\includegraphics[bb=40 350 530 770,clip=]{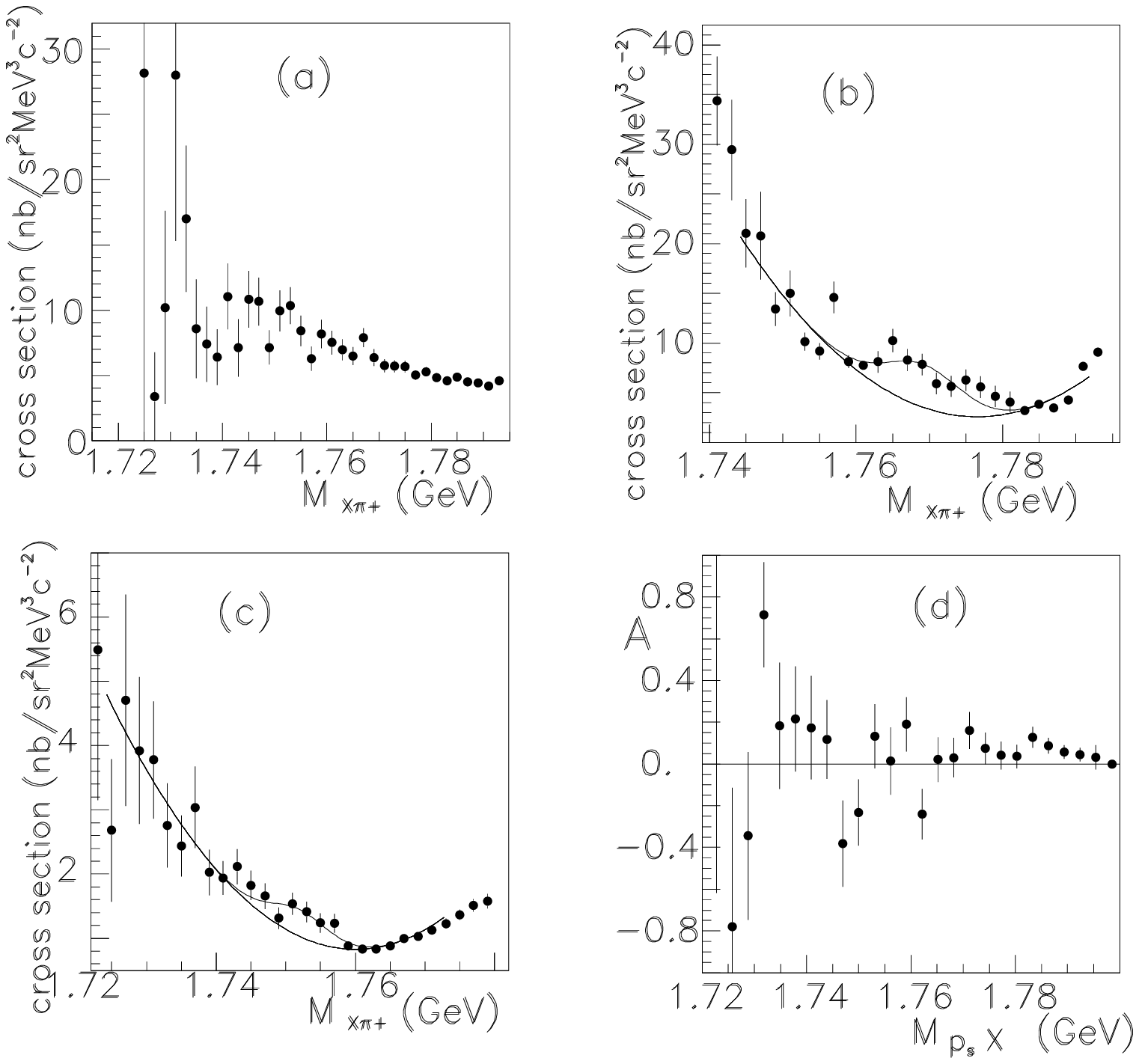}}
\caption{Inserts (a), (b), and (c) show the differential
cross sections of the reaction
p~p~$\to~p~\pi^{+}$~X, at T$_{p}$=2.1 GeV, in the laboratory system, at
$\theta$=0.7$^{0}$, 3$^{0}$, and 9$^{0}$ respectively. Insert (d) shows
the analyzing power of the p~p~$\to$~p$_{f}$~p$_{s}$~X reaction,
at T$_{p}$=2.1 GeV, $\theta$=3$^{0}$.
Values are listed in Table~1.}
\label{fig6}
\end{figure}
\end{center}

\subsection{Simulation}
\hspace*{0.4cm}The experiment was simulated in order to
check several aspects of the detection system, and to evaluate the data
normalization for inefficiencies and
acceptances. Differential cross sections could then be evaluated.
The correction function was smooth for all cross section variables.
The simulation describes the detector and the
spectrometer magnetic field properties and generates events for a
given missing mass distribution and a given angular cross section
distribution over the spectrometer horizontal angular range. The code
was also used to study the
consequences of particles scattered by the target in the vertical plane at
angles $50 \le \verb/|/\phi \verb/|/ \le 80 $ mrd.
No bias corresponding to the appearance of narrow structures was observed
in the simulation. Figure~2 illustrates the comparison between the experiment
and the simulation for the pp$\rightarrow$p$\pi^{+}$n reaction. Parts 2(a)
and (c), showing the pion proton momenta correlation, are in excellent
agreement, as are the missing mass distributions shown in parts 2(b) and (d).
The variation of the cross section inside the spectrometer solid angle
($\pm 3^{0}$) in both horizontal and vertical planes does not reproduce
exactly the experimental cross section angular dependance inside the
spectrometer solid angle. This is not relevant for the present
discussion.\\
\hspace*{0.4cm}We observe that the simulation reproduces the position and
the width of the neutron missing
mass peak. Indeed in the range 930$\le$~M~$\le$~950~MeV, the $\sigma$ of
the neutron peak in the data (simulation) is equal to 3.4 (3.2)~MeV.
\subsection{Controls}
Many checks were performed to establish whether or not the observed
structures are results of genuine physical phenomena or of experimental
artefacts. In the following the various results of the tests will be
outlined briefly but can be examined in detail in a previous paper 
\cite{bor5}.\\
\hspace*{0.4cm}
Different cuts on particle momenta, as well as on emission angles, were
performed over a mass range where structures larger than those
discussed here were observed  \cite{bor5}. The effect of these cuts,
is to reduce the statistics (see figures 11, 12, and 13 of reference 
\cite{bor5}) and hamper the
interpretation of the data. Nonetheless, the three
structures under discussion there, remained in place.\\
\hspace*{0.4cm}
It was also shown in figure 14 of reference \cite{bor5} that the
structures did not
depend on the spin state of the incident beam. In figure 15 of the same
paper, it is also shown that the structures were not
present in accidental coincidences.\\
\hspace*{0.4cm}
The empty target measurements performed as a control and for background
subtraction, showed no target contamination (see figure 16 of
reference \cite{bor5}).  It is also concluded that, although
the measurements were performed at small angles, there was no evidence
of any hot area of incident beam which could have been
scattered by any mechanical piece at the entrance of the spectrometer.\\
\hspace*{0.4cm}
A detailed discussion was presented that allowed to eliminate
contamination from the possible
effect of particles emitted outside the spectrometer solid angle but
subsequently slowed
down (figures 18 and 11 of the same paper).\\
\section{Results}
\hspace*{0.4cm}At $T_{p}$=2.1 GeV this experiment was performed at three
spectrometer angles:
$\theta_{lab.}$=0.7$^{0}$, 3$^{0}$, and 9$^{0}$.
Peaks were tentatively extracted using low order
polynomials for the background and a Gaussian peak for the structure.
The channels defining the polynomial and Gaussian peak ranges were given,
then all parameters were allowed to vary freely.
The following relation was used for the standard deviation (S.D.)
evaluation~:
\begin{equation}
S.D. = [\Sigma_{i}(N_{Ti} - N_{Bi})/\Delta\sigma_{i}^{2}]/[{\Sigma_{i}
(1/\Delta\sigma_{i}^{2}})]^{1/2}
\end{equation}
where N$_{Ti}$ (N$_{Bi}$) corresponds to the total (background) number of
events for the channel i,
and $\Delta\sigma_{i}^{2}$ = $\Delta\sigma_{Ti}^{2}$ +
$\Delta\sigma_{Bi}^{2}$~$\approx$~2~$\Delta\sigma_{Ti}^{2}$.
The quantitative information is summarized in Table 1 where we observe
small fluctuations of the structure's masses. The source of these
fluctuations, of the order of $\Delta$M/M $\approx 10^{-3}$ are attributed
to three experimental limits. The first limit comes from the mass calibration.
Since there is no known reference peak within the experimental acceptance, a
high precision mass calibration control cannot be carried out. The second
limit is statistic: in order to have enough statistical precision the data
are integrated over a large spectrometer angle ($\pm 3^{0} _{lab.}$). The
last source of small fluctuations, can be attributed to using a single set
of reconstruction parameters over the whole experimental range.\\
\subsection{The p~p~$\to$~p$_{s}$~p$_{f}$~X reaction}
\hspace*{0.4cm}The measured differential mass spectra are shown in
figure~4. Structures close to M=1747~MeV (1772~MeV) are observed with large
S.D. in inserts (a), (b), and (d) ((b) and (c)).
At $\theta$=0.7$^{0}$ (insert (a)), no peak was extracted close to
M=1772 MeV. In this case, the physical background being larger than
at $\theta$=3$^{0}$
(insert (b)) and the statistics lower, a possible
peak at $\theta$=0.7$^{0}$ could have a cross section similar
to the one extracted at $\theta$=3$^{0}$. Well defined peaks are observed
at $\theta$=3$^{0}$ with their cross sections decreasing at
$\theta$=9$^{0}$.\\
\hspace*{.4cm}
The rapidly increasing shape of the invariant mass data $M_{p_{f} X}$
does not allow any peak extraction at $\theta$=0.7$^{0}$ or
$\theta$=3$^{0}$. However, at $\theta$=9$^{0}$, a structure at
M=1747~MeV was
extracted (insert (d)). Here the cut used to define the upper branch is
$p_{p_{s} X}\ge$~0.97~GeV/c.\\
\hspace*{.4cm} In Figure~4(d) a single data point close to M=1758~MeV,
is higher than its neighbours by
several S.D. For this reaction and angle, the acceptance spectrum
has a peak at this mass. A small shift in mass between the
data and the momentum acceptance used for the normalization of
the data creates this effect.
A careful control was performed in order to check that the
physical peaks could not arise from such artefacts.\\
\hspace*{.4cm}
The experiment was performed using polarized incident protons and
the analyzing power (figure~6(d)) shows several discontinuities. These
effects are not statistically significant, but they are
shown since they appear at the same masses where statistically
significant structures are observed in the cross sections.
In this case the data are integrated over
3 MeV bins as opposed to 2 MeV bins as in figure~4(b).
\subsection{The p~p~$\to$~p~$\pi^{+}$~X reaction}
For this reaction both root solutions for M$_{\pi^{+} X}$ were separated
by a software cut on the proton momentum.  Figure 5
shows the cross sections of the upper branch where $p_{p}\ge$1~GeV/c.
Figures 6(a), (b), and (c) show the cross sections of
the lower branch where $p_{p}\le$0.95~GeV/c.
\subsubsection{The upper branch}
The statistics in the upper branch are higher than in the lower
branch. However, part of the upper branch statistics included data from the
tail of the neutron missing mass, particularly at forward angles.
Supplementary software cuts on M$_{X}$ were therefore applied in order to
suppress the neutron peak contamination.
In figures~5~(a) and (b) the invariant mass spectra obtained at
$\theta$=0.7$^{0}$ are shown. Cuts on $M_{X}$ at $M_{X}\ge$1.15~GeV and
$M_{X}\ge$1.3~GeV respectively were applied. When going from 5(a) (2 MeV
binning) to (b) (4 MeV binning), the differential
cross sections decrease but the peak's heights are enhanced.
The 1747~MeV structure is observed in both binnings. The 1773~MeV structure
is observed in figures~5(b), (c), and (d), but is pratically absent in
figure~5(a). A not excluded lower background in figure~5(a) below M=1772~MeV
will allow a structure extraction at such mass.\\
\hspace*{.4cm} We recall that
all data are normalized by the momentum phase space
$\Delta$p$_{p}\times\Delta$p$_{\pi}$. Consequently the relative comparison
of different cross sections cannot be made when different cuts are applied.
In figures~5(c) and (d) the \\
\begin{table}
\caption{Properties of the tentatively extracted narrow
peaks in the 1720-1790~MeV baryonic mass region as observed in
pp$\rightarrow$~p$\pi^{+}$X and pp$\rightarrow$~ppX reactions at
T$_{p}$=2100~MeV.}
\label{Table 1}
\end{table}
\vspace*{-0.3cm}
\begin{center}
\begin{table}
\begin{tabular}[h]{c c c c c c c c c c c}
\hline
M&Fig&Mass&obs.&width&SD&reaction&$\theta$&cuts\\
(MeV)& &(MeV) & & & & & &\\
\hline
1747&4(a)&1745.1&M$_{p_{s}X}$&1.3&11.7&$pp\rightarrow ppX$&0.7$^{0}$&upper
branch\\
&4(b)&1745.9&M$_{p_{s}X}$&2.5&9.1&$pp\rightarrow~ppX$&3$^{0}$&upper branch\\
&4(d)&1747.0&M$_{p_{f}X}$&4.5&3.4&$pp\rightarrow~ppX$&9$^{0}$&upper branch\\
&5(a)&1746.6&M$_{\pi{+}X}$&1.5&3.6&$pp\rightarrow~p\pi^{+}X$&0.7$^{0}$&(see
text)\\
&5(b)&1747.9&M$_{\pi{+}X}$&2.1&2.5&$pp\rightarrow~p\pi^{+}X$&0.7$^{0}$&(see
text)\\
&6(c)&1751.9&M$_{\pi{+}X}$&4.2&4.3&$pp\rightarrow p\pi^{+}X$&9$^{0}$&lower
branch\\
\hline
1772&4(b)&1773.3&M$_{p_{s}X}$&2.9&9.4&$pp\rightarrow ppX$&3$^{0}$&upper
branch\\
&4(c)&1770.6&M$_{p_{s}X}$&2.0&3.8 &$pp\rightarrow ppX$&9$^{0}$&upper branch\\
&5(b)&1771.5&M$_{\pi{+}X}$&2.1&4.7
&$pp\rightarrow~p\pi^{+}X$&0.7$^{0}$&(see text)\\
&5(c)&1773.5&M$_{\pi{+}X}$&4.9&10.5&$pp\rightarrow~p\pi^{+}X$&3$^{0}$&
upper branch\\
&5(d)&1773.6&M$_{\pi{+}X}$&3.4&5.8 &$pp\rightarrow~p\pi^{+}X$&9$^{0}$&
upper branch\\
&6(b)&1769.1&M$_{\pi{+}X}$&5.3&7.8&$pp\rightarrow~p\pi^{+}X$&3$^{0}$&
lower branch\\
\hline
& 6(d)& & M$_{p_{s}X}$& & & $pp\rightarrow ppX$&3$^{0}$&lower branch\\
\hline
\end{tabular}
\end{table}
\end{center}
peaks extracted at $\theta$~=~3$^{0}$ (M$_{X}\ge$1.15~GeV) and
$\theta$~=~9$^{0}$ (M$_{X}\ge$0.98~GeV) are shown respectively.\\
\subsubsection{The lower branch}
In figures~6(a), (b), and (c) the differential cross sections
of the $M_{X\pi^{+}}$ missing mass are shown for the
p~p~$\to$~p~$\pi^{+}~$X reaction. Low branch events are defined by the
condition that $p_{p}\le$~0.95~GeV/c.
The data are shown for $\theta$=0.7$^{0}$, 3$^{0}$, and 9$^{0}$
respectively.
The error bars are too large at $\theta=0.7^{0}$ to allow a peak to be
extracted
around M=1747 MeV. In all cases, the available statistics are poor.
\subsection{Differential cross sections}
The differential cross sections were obtained for all peaks extracted. Since
they are not well defined they
are not included in Table~1. Their variation versus the spectrometer angle is
only meaningful when all cuts are the same. In this case, we observe
that the cross section decreases by a factor $\approx$~2.5 when the
spectrometer angle increases from
$\theta$=0.7$^{0}$ up to
$\theta$=3$^{0}$ and decreases by a further factor $\approx$~7 as the
spectrometer angle increases from
$\theta$=3$^{0}$ up to $\theta$=9$^{0}$.
\subsection{Other spectra - not presented.}
\hspace*{4.mm}In Table~2, a summary of the overall situation is given. In
this table there are references to spectra shown in this paper as well as
other spectra, which are not shown since there were no structures
observed.\\
\hspace*{4.mm}As a complement to figures 4, 5, and 6, there was also a study
of the forward kinematical branch of the p~p~$\to$~p~p~X reaction at
$\theta$~=~0.7$^{0}$ and $\theta$~=~3$^{0}$. No peak or structure was
observed in the M$_{p_{f}}$ missing mass spectra.\\
\hspace*{4.mm}
The lower branch of the p~p~$\to$~p~p~X reaction was also studied. Both the
M$_{p_{s}X}$ and M$_{p_{f}X}$ missing masses were displayed along with
their differential cross sections. The spectra had similar shapes, both
having low statistics and a rapid rise in the cross section as a function of
the missing mass. The statistics were only sufficient for a more detailed
study for the M$_{p_{s}X}$ data, inside a very narrow invariant mass
range, between M=1782~MeV and M=1791~MeV.\\
\hspace*{4.mm}
It is quite natural to observe higher statistics in the $M_{p_{s}X}$ upper
branch, as compared to the lower branch. This is because the selection is
made on the fast proton and there are more events in the upper branch of
the fast proton
than in the lower branch. Symmetrically, we would expect
to have the $M_{p_{f}X}$ lower branch
favoured relative to the upper branch, but such a
comparison must be put alongside the observation that, with our
experimental kinematical conditions, there are naturally many more events
in the upper branch than in the lower branch.
For example, if we study the kinematics of the two particle
reaction: p~p~$\to~$p$_{3}$N$_{4X}$ with the mass M(N$_{4X}$)=1760 MeV, at
$\theta=1^{0}$, the
forward solution corresponds to $p_{3}$=1310 MeV/c, and the momentum
transfer q=4.0~fm$^{-1}$, whereas the backward solution corresponds to
$p_{3}$=720 MeV/c, and the momentum transfer q=6.1 fm$^{-1}$. This larger
value of q also explains the difficulty in observing small structures. The
spectrum of $M_{\pi^{+}X}$ at $\theta$=0.7$^{0}$, from
the $pp\rightarrow p\pi^{+}X$ reaction (lower branch), is not shown. Here
the statistics are close to 300 events/MeV at $M_{\pi^{+}X}$=1970 MeV and
are even lower for smaller masses. The same argument as before concerning
the transfer momenta values between forward and backward solutions holds
here.\\
\subsection{Comparison Between Data And Simulation}
\hspace*{4.mm}
Two structures were extracted at 1747 MeV and 1772 MeV.
All the peaks - except one - were extracted with S.D.$\ge$3.3.
They are shown graphically in figures 4, 5, and 6, and numerically
in Table~1. \\
\hspace*{0.4cm}In figure~7 the comparison between data
and simulation is made
for the p~p~$\rightarrow$~p~p~X reaction at $T_{p}$=2.1~GeV and
$\theta$=3$^{0}$ (upper branch).
It corresponds to the data shown in Figure~4(b) (see also Table~1).
Apart from the extremities, we observe a similar shape with a small excess
in narrow ranges of data. These excesses occur at the same masses where
the possibility of a peak was observed.\\
\begin{center}
\begin{figure}
\vspace{-0.1cm}
\includegraphics[bb=0 260 530 530,clip,scale=0.8]{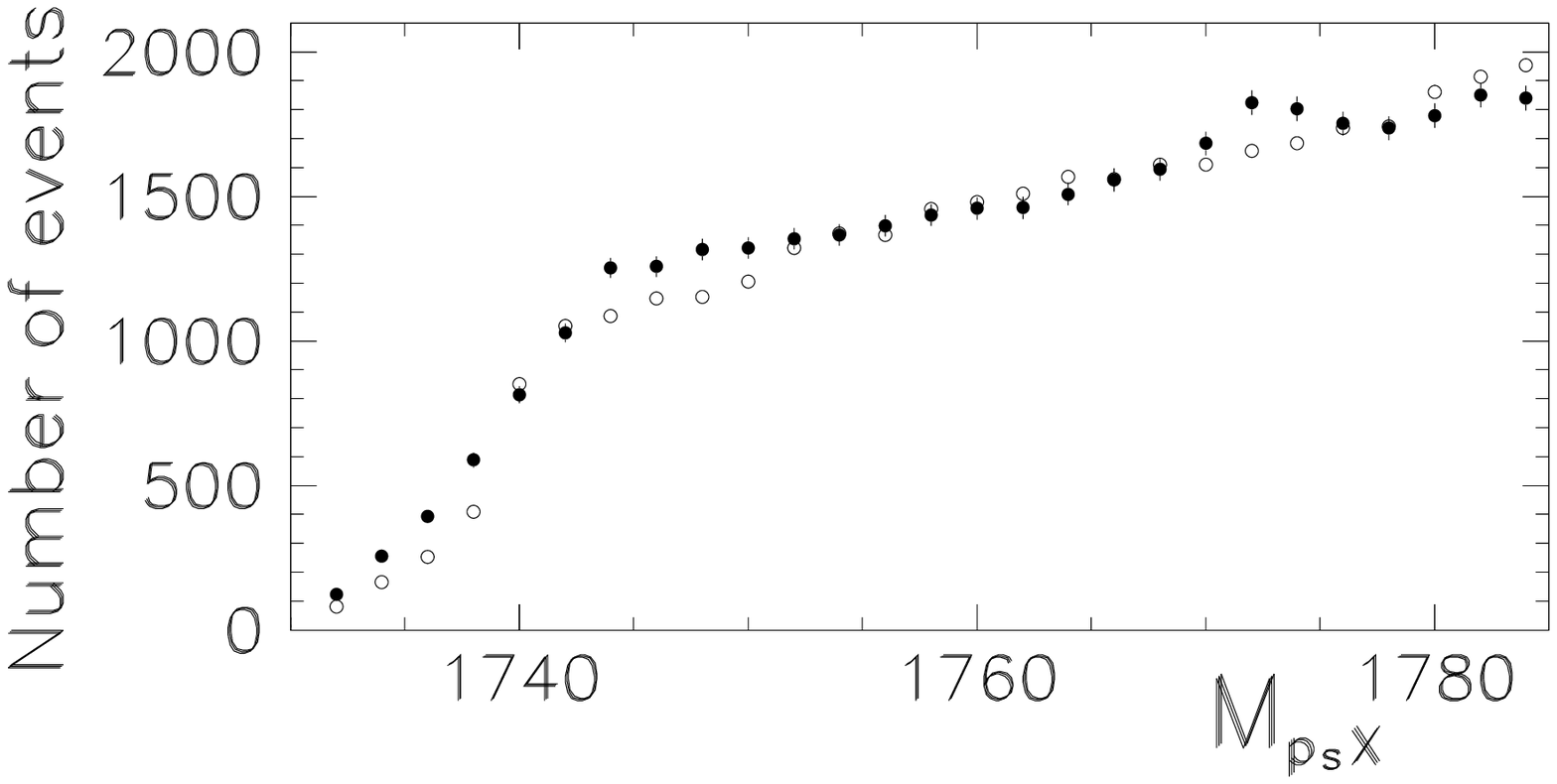}
\caption{Comparison between data (full points) and simulation
(empty points) for the
pp$\rightarrow$ppX reaction at $T_{p}$=2.1~GeV and $\theta$=3$^{0}$
(upper branch). The experimental data are the same as those shown in
figure~4(a).}
\label{fig7}
\end{figure}
\end{center}
\vspace{-.9cm}
\section{Discussion}
\hspace*{0.4cm} Since the two structures extracted at M=1747 MeV and
M=1772 MeV are both narrow and are excited at stable masses
regardless of the reaction or the angle, then it is not unreasonable
to suggest that they may be exotic baryons.
Their widths are narrow, and therefore the structures may be
associated with genuine baryonic states as opposed to being from some
dynamical rescattering effects between final state particles. The mean
extracted widths for $\theta$=0.7$^{0}$, 3$^{0}$, and
9$^{0}$ are $\sigma$=1.8 MeV, 3.9 MeV, and 3.1 MeV respectively.
Although the precision is low, these values agree with the experimentally
simulated resolutions.\\
In the p~p~$\rightarrow$~p~p~X reaction, the missing mass range
640$\le M_{X}\le$840 MeV, corresponds mainly to two-pion phase space
\cite{yonn} and $\rho^{0}$ and $\omega$ mesons.
Therefore, these structures may couple
to $\Delta\pi$, p$\pi\pi$, p$\rho^{0}$, or p$\omega$ final states.\\
\hspace*{0.4cm}The coupling of these potentially new narrow baryons to
N$\pi$ cannot be easily studied. Indeed, if the
pp$\rightarrow$p$\pi^{+}$n reaction allows us to study such a coupling,
the software cuts due to the experimental acceptance and the need to
eliminate the range
1746~$\le$ M$_{\pi^{+}n}\le$~1763 MeV,
which corresponds to a region of lost events where
p$_{p}\approx$~p$_{\pi^{+}}$, prevents us from getting a clear answer.
Nonetheless, our results do not contradict a twenty-year old theoretical
statement \cite{koni} that the missing baryons couple weakly to N$\pi$.\\
\hspace*{0.4cm}Several broad baryons, in the
1700$\le$M$\le$1800 MeV mass range, quoted by PDG \cite{pdg} with at least
two $\star$'s, are N(1700)D$_{13}$,
N(1710)P$_{11}$, N(1720)P$_{13}$ and $\Delta$(1700)D$_{33}$. They all have
full widths $\ge$100 MeV and a predominant decay mode to N$\pi\pi$.
Interferences between broad baryons located on both sides of the range
1700$\le$M$\le$1800~MeV may be relevant. A simulation was performed using
Breit Wigner shapes with physical masses and widths given in PDG and
arbitrary amplitudes (a). Figure~8 shows the results of the
simulation for three different spin and isospin states. The masses, widths
and arbitrary amplitudes are presented in Table~3.
In Figure~8 the light lines
correspond to the same calculation, but in this case the amplitude of
the second baryon is taken as being
negative. We observe that in all cases, a narrow, resonance-like structure
cannot be created by an interference
between broad baryons. Their main effect is to shift,
by a small amount, the position of the maximum, which remains broad anyway.
So, the very different masses and widths arising from possible interferences
between broad baryons, and those observed in our data, preclude any common
identification.\\
\hspace*{0.4cm} The same baryonic mass region
was previously studied for other purposes with a lesser resolution and (or)
weaker statistics and larger data bins. However, some recent
results show some narrow structures, although these are not pointed out
by the authors (usually due to
statistical reasoning). For example, in the $ep\to~e'p\eta$ cross sections
measured at CEBAF \cite{thom}, a sharp structure
was observed at W$\approx$1.7 GeV
and was shown to originate from an interference between S and P waves,
in terms of known resonances. Its width is about one order of
magnitude larger than the width of the structures
presented in this paper, and so in this case, the source of the peak is well
understood. In the raw asymmetry data measured with
CLAS at JLAB at $T_{e}$=2.6 GeV in an inclusive
$ \rm\vec{e}~\rm\vec{NH_{3}}\to$~eX scattering experiment \cite{burk}, a
peak at\\
\vspace{-0.2cm}
\begin{table}[h]
\caption{\label{Table 2}Description of all the different spectra obtained
from the analysis.}
\label{tab2}
\end{table}
\vspace{-0.3cm}
\begin{center}
\begin{table}[b]
\begin{tabular}{c c c c}
reaction&selection&variable&figures or comments\\
\hline
$pp\rightarrow ppX$&upper branch&
M$_{p_{s}X}$&0.7$^{0}$: Figure~4(a),\hspace*{2.mm}3$^{0}$: Figure~4(b)\\
& & &9$^{0}$: Figure~4(c)\\
& &M$_{p_{f}X}$& 9$^{0}$: Figure~4(d)\\
&lower branch&M$_{p_{s}X}$&poor statistics and rapidly rising spectra\\
&          &M$_{p_{f}X}$&poor statistics and rapidly rising spectra\\ 
\hline
$pp\rightarrow p\pi^{+}$X&upper branch&M$_{\pi^{+}X}$&0.7$^{0}$:
Figure~5(a),\hspace*{2.mm}3$^{0}$: Figure~5(c)\\
& & &9$^{0}$: Figure~5(d)\\
&lower branch&M$_{\pi^{+}X}$ &3$^{0}$: Figure~6(b),\hspace*{2.mm}9$^{0}$:
Figure~6(c)\\
\end{tabular}
\end{table}
\end{center}

\begin{center}
\begin{figure}
\includegraphics[bb=5 5 540 540,clip,scale=0.6]{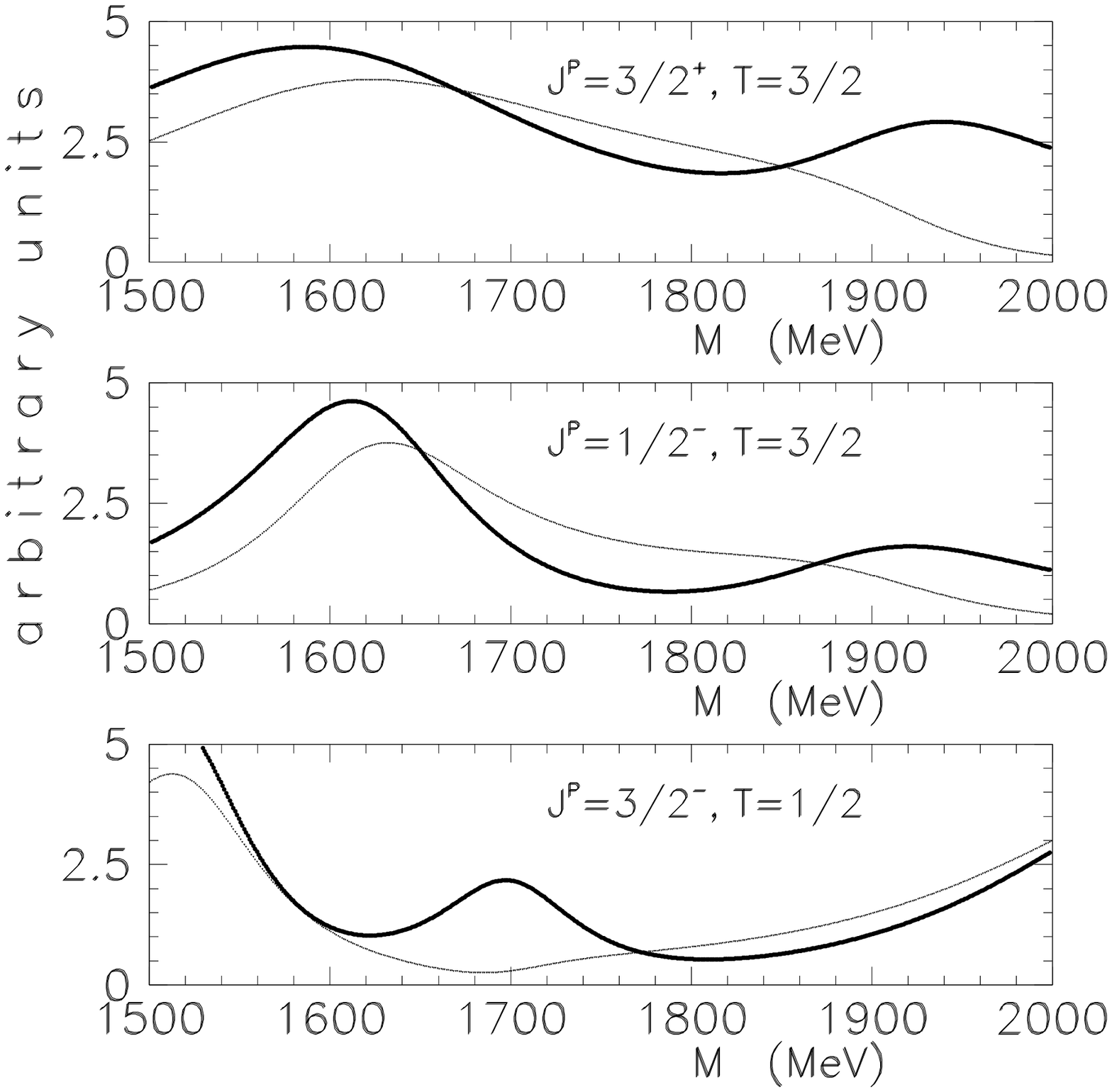}
\caption{Search for possible narrow structures created by
interferences between broad baryons quoted by PDG [6]. Table~3 lists the
masses, widths and amplitudes used.}
\label{fig8}
\end{figure}
\end{center}
 M$\approx$1780~MeV ($\sigma\approx$10 MeV) was also observed.
This peak was not discussed by the authors,
although its S.D.~$\approx$~3~MeV.
Many channel data, ep$\to$eX and ep$\to$e$\Delta^{++}\pi^{-}$
were recorded simultaneously using CLAS at JLAB \cite{burk1}. In these
data, the
rather large binning precludes the observation of eventual narrow
structures.\\
\hspace*{0.4cm} The various models of the baryon spectra predicted in the
symmetric $\mid q^{3}>$ model
were described in a review article by Capstick and Roberts
\cite{caps}. This article contains an important list of references.
In the mass range close to the one discussed in this paper, several
baryons\\
\begin{table}
\caption{Description of the broad baryon's used for the
simulation of possible narrow bumps through interferences. Masses and
widths are in MeV.}
\label{tab3}
\end{table}
\begin{center}
\begin{table}
\begin{tabular}{c c c c c c c c c c c}
J$^P$&T&M$_{1}$&($\Gamma _{1}$)& a$_{1}$&M$_{2}$&$\Gamma _{2}$)&a$_{2}$
&M$_{3}$&($\Gamma _{3}$)& a$_{3}$\\
\hline
3/2$^+$&3/2&1600&(350)&2&1920&(200)&1&&&\\
1/2$^-$&3/2&1620&(150)&2&1900&(200)&1&&&\\
3/2$^-$&1/2&1520&(120)&2&1700&(100)&1&2080&(300)&2\\
\end{tabular}
\end{table}
\end{center}
\vspace*{0.cm}
 were predicted. Many of them may be associated with known PDG
baryonic resonances with the same quantum numbers. However, some have no
experimental equivalence if we only associate each known PDG resonance
with a single calculated baryonic mass.
The non-relativistic QCD model \cite{isgu}
predicted several states, all of which have a broad PDG baryon counterpart.
In the quark model using chromodynamics \cite{koni1}  
\cite{koni2} \cite{koni3}, one level N(1870)P$_{13}$
has no PDG counterpart.
The relativized quark model \cite{caps1} \cite{caps2} predicts one baryon
$\Delta\frac{1}{2}^{+}$ at 1835 MeV without having any PDG counterpart,
but with a large partial width in the $\Delta \pi$ channel.
Within the extended Goldstone-Boson-Exchange Constituent Quark Model
\cite{gloz} \cite{wage}, four levels are found between M$\approx$1700 and
1780 MeV, all of which have a known PDG baryon counterpart if we accept
identifications
with a difference as large as 100 MeV between calculated and experimental
masses. The collective algebraic model \cite{bijk} \cite{bijk1}, predicted
several levels,
among which four have no  PDG counterpart, namely one N$\frac{1}{2}^{+}$
state at M$\approx$1725, two N$\frac{3}{2}^{+}$ states
at  M$\approx$1740 MeV and M$\approx$1810
MeV, and one $\Delta\frac{3}{2}^{+}$ baryon
at M$\approx$1910 MeV.
There are no missing baryons in the mass range considered in the
semi-relativistic flux-tube model \cite{stas} \cite{stas1}.
There are no missing states for masses lower than 2.4 GeV within the framework
of the relativistic covariant quark model with instanton-induced quark
forces \cite{lori}.\\
\hspace*{0.4cm}Among this multitude of predicted baryons, only
those having a small calculated width can be compared to our narrow
structures. Our pp$\rightarrow$p$\pi^{+}$X reaction selects mainly the
$\Delta\pi$ invariant mass, whereas the pp$\rightarrow$ppX
reaction selects $\Delta\pi$, N$\pi\pi$, and N$\pi\pi\pi$ invariant masses.
Therefore, the small experimental widths can be close to total widths.
Although the calculated partial widths of the missing baryons are sometimes
small, even in the $\Delta \pi$ channel, it is unlikely that the total widths
would remain so small for $\mid q^{3}>$ baryons.
\section{Conclusion}
\hspace*{0.4cm}Due to the good resolution of our experimental set-up,
previously unobserved narrow structures were identified.
An indication for narrow baryonic structures at
M=1747~MeV and 1772~MeV were observed in the invariant
mass spectra M$_{pX}$ (M$_{\pi^{+}X}$)
of the  pp$\rightarrow$ppX (pp$\rightarrow$p$\pi^{+}$X) reaction.
These structures were extracted from several spectra, each with a large
number of standard deviations. A careful
comparison of our data to simulated spectra allow to conclude
that these structures are not simply experimental artefacts. Their
small widths exclude the possibility of associating these
structures with interference effects between final state particles.
Considering the stability of the masses extracted from various reactions at
various angles, and with different filters applied, it is reasonable to
suppose that these structures are
genuine baryonic states. Their masses and widths are compared to the
masses and widths of the missing
baryon model calculations, and it is concluded that such an identification
is unlikely. These possible
baryons could be more exotic than the three-quark
broad baryons. The name `exotic'' is usually associated with meson
production \cite{seth}, where angular distributions are measured and the
extracted quantum numbers forbid simple $q\bar{q}$ configurations.
In the present work, no spin assignment could be
made since no angular distributions are available. The name `exotic''
is supported by the
small observed widths and the impossibility to associate them with
any classical
$q^{3}$ configuration.\\
\hspace*{0.4cm}The small spacing $\approx$~25 MeV
between the observed masses suggests that many narrow and
weakly excited
states could exist within the baryonic spectra, and are superimposed
on the classical broad baryons.
The small widths of the states
observed in the present work suggest that the  wave functions of these
possible baryons could be more
complicated than the wave functions of $\mid q^{3}>$ baryons.

\end{document}